\documentclass[reprint,twocolumn,superscriptaddress,showpacs,nofootinbib,notitlepage,preprintnumbers,aps,prd,floatfix]{revtex4-2}
\usepackage[utf8]{inputenc}
\usepackage{graphicx}
\usepackage{latexsym,amsmath,amssymb,lmodern,float,url}
\usepackage{natbib}
\usepackage{color}
\usepackage{microtype}
\usepackage{bbold}
\usepackage{slashed}
\usepackage{multirow}
\usepackage{tikz}
\usepackage{xcolor}
\usepackage{mathrsfs}  
\usetikzlibrary{shapes}
\usepackage{adjustbox}

\usepackage[colorlinks=true,backref=false, linktocpage=true,
citecolor=blue,urlcolor=blue,linkcolor=blue,pdfpagemode=UseOutlines]{hyperref}

\hypersetup{%
 bookmarksnumbered=true,
 pdftitle = {},
 pdfsubject = {},
 pdfauthor = {},
 pdfkeywords = {}
}

\let\Re\undefined

\newcommand{\ds}{\mathbb{V}}

\DeclareMathOperator{\Tr}{Tr}
\DeclareMathOperator{\tr}{Tr}
\DeclareMathOperator{\Re}{Re}

\def \matrix #1 {\left(\begin{array}{cc} #1 \end{array}\right)}
\def \Tr {\mathop{\rm Tr}\nolimits}
\def \tr {\mathop{\rm tr}\nolimits}

\def \Re {\mathop{\rm Re}\nolimits}

\def \e  {\mathop{\rm e}\nolimits}

\def\II{\hbox{{1}\kern-.25em\hbox{l}}}

\definecolor{vermillion}{RGB}{213,94,0}
\definecolor{cblue}{RGB}{0,114,178}
\definecolor{corange}{RGB}{230,159,0}
\definecolor{blgreen}{RGB}{0,158,115}
\definecolor{repurple}{RGB}{204,121,167}

\definecolor{mycolor}{RGB}{0,255,0}
\newcommand{\redsq}{\raisebox{0.5pt}{\tikz{\node[fill,scale=0.5,regular polygon, regular polygon sides=4,fill=vermillion](){};}}}
\newcommand{\blcir}{\raisebox{0.5pt}{\tikz{\node[fill,scale=0.5,circle,fill=cblue](){};}}}
\newcommand{\ortri}{\raisebox{0.7pt}{\tikz{\node[fill,scale=0.4,regular polygon, regular polygon sides=3,fill=corange,rotate=0](){};}}}
\newcommand{\grtri}{\raisebox{0.7pt}{\tikz{\node[fill,scale=0.4,regular polygon, regular polygon sides=3,fill=blgreen,rotate=180](){};}}}
\newcommand{\pudia}{\raisebox{0.7pt}{\tikz{\node[fill,scale=0.4,regular polygon, regular polygon sides=4,fill=repurple,rotate=45](){};}}}

\begin{document}
 \preprint{FERMILAB-PUB-21-496-T, TUM-HEP-1368/21}
\title{Gluon Digitization via Character Expansion for Quantum Computers}
\author{Yao Ji}
\email{yao.ji@tum.de}
\affiliation{Physik Department T31, James-Franck-Stra\ss e 1, Technische Universit\"at M\"unchen, D-85748 Garching, Germany}
\author{Henry Lamm}
\email{hlamm@fnal.gov}
\affiliation{Fermi National Accelerator Laboratory, Batavia, Illinois, 60510, USA}
\author{Shuchen Zhu}
\email{sz424@georgetown.edu} 
\affiliation{Department of Computer Science, Georgetown University, Washington, DC 20057, USA}
\date{\today}
\collaboration{NuQS Collaboration}
\begin{abstract}
Efficient digitization is required for quantum simulations of gauge theories. Schemes based on discrete subgroups use a smaller, fixed number of qubits at the cost of systematic errors. We systematize this approach by deriving the single plaquette action through matching the continuous group action to that of a discrete one via group character expansions modulo the field fluctuation contributions. We accompany this scheme by simulations of pure gauge over the largest discrete crystal-like subgroup of $SU(3)$ up to the fifth-order 
in the coupling constant. 
\end{abstract}
\maketitle
\section{Introduction}
Quantum computing has the potential to dramatically  advance our understanding of quantum field theory~~\cite{Feynman:1981tf,Jordan:2017lea,Banuls:2019bmf} -- especially for the dynamics of QCD~\cite{Lamm:2019uyc,Kreshchuk:2020dla,Echevarria:2020wct}.
Performing such simulations will require a large, albeit at present unknown,  scale of the quantum computers in terms of memory and circuit depths~\cite{Kan:2021xfc}. This scale is certainly beyond the current, noisy devices.  
Thus, it is therefore essential to explore the possibilities of reducing the requirement of quantum resources so that any near-future field theory simulations on a quantum computer become feasible.
Such investigations could not only help to determine the advantages of various quantum algorithms in different circumstances, but also be beneficial in the future when quantum computing becomes commonplace by providing general frameworks for cost-effective QCD simulations.

For real-time QCD simulations, large quantum resources are allocated to digitize the gluon fields due to their bosonic nature and thus unbounded Hilbert space. Many approaches exist with different and presently poorly-understood costs. Prominent proposals for digitization~\cite{digi_loi} can be broadly classified into: Casimir dynamics~\cite{Zohar:2012ay,Zohar:2012xf,Zohar:2013zla,Zohar:2014qma,Zohar:2015hwa,Zohar:2016iic,Klco:2019evd,Ciavarella:2021nmj} potentially with auxiliary fields~\cite{Bender:2018rdp}, conformal truncation~\cite{Liu:2020eoa}, discrete groups~\cite{Hackett:2018cel,Alexandru:2019nsa,Yamamoto:2020eqi,Ji:2020kjk,Haase:2020kaj}, dual variables~\cite{PhysRevD.99.114507,Bazavov:2015kka,Zhang:2018ufj,Unmuth-Yockey:2018ugm,Unmuth-Yockey:2018xak}, light-front kinematics~\cite{Kreshchuk:2020dla,Kreshchuk:2020aiq}, loop-string-hadron formulation~\cite{Raychowdhury:2018osk,Raychowdhury:2019iki,Davoudi:2020yln}, meshes and subsets~\cite{Hackett:2018cel,Hartung:2022hoz}, quantum link models~\cite{Wiese:2014rla,Luo:2019vmi,Brower:2020huh,Mathis:2020fuo}, and qubit regularization~\cite{Singh:2019jog,Singh:2019uwd,Buser:2020uzs}.  Each approximation reduces symmetries -- either explicitly or through finite-truncations~\cite{Zohar:2013zla}. Thus, one must proceed with caution as the regulated theory may not have the original theory as its continuum limit~\cite{Hasenfratz:2001iz,Caracciolo:2001jd,Hasenfratz:2000hd,PhysRevE.57.111,PhysRevE.94.022134,article}.  In this spirit, a number of recent works have attempted to quantify the truncation needed to achieve fixed accuracy for a lattice simulation~\cite{Davoudi:2020yln,Shaw:2020udc,Kan:2021xfc,Tong:2021rfv}. 

Studies of the quantum simulations of lattice field theories typically use the Hamiltonian formalism, with the Kogut-Susskind Hamiltonian~\cite{PhysRevD.11.395} being the most investigated for gauge theories.  In this work, we will instead work in the action formalism.  This allows us to construct modified actions that can be studied nonperturbatively on classical computers today.  For the eventual simulation on quantum computers, one can derive the modified Hamiltonian straightforwardly via the transfer matrix~\cite{Creutz:1984mg,Carena:2021ltu,Luo:1998dx}.

One strategy pursued in the early days of lattice QCD involved approximating the gauge group $SU(3)$ by its largest crystal-like subgroup $\ds$~\cite{Flyvbjerg:1984dj, Flyvbjerg:1984ji}, thus maintaining a remnant of the gauge symmetry of the parent group.  Depending on the action chosen, the accuracy of the approximation varies greatly. The first studies considered merely replacing the continuous group $G$ by its discrete subgroup $H$ in the Wilson gauge action. The viability of this approximation was studied in detail for $U(1)$~\cite{Creutz:1979zg,Creutz:1982dn}, and $SU(N)$~\cite{Bhanot:1981xp,Petcher:1980cq,Bhanot:1981pj}, with the inclusion of  fermions~\cite{Weingarten:1980hx,Weingarten:1981jy}.
These studies met with mixed success.  In order to improve the agreement between lattice observables for $H$ and $G$, modified actions in $H$ were considered, which generally lead to vastly improved results~\cite{Alexandru:2019nsa,Alexandru:2021jpm}.  The drawback to these ad-hoc actions is that they were found empirically through substantial classical simulations, and thus a theoretical and systematic understanding of them is lacking.

The resulting discrepancy from such approximations can be analyzed systematically through expansions of various parameters in the same spirit of the modern effective field theory (EFT) approach with a matching procedure.
So far, two parameters have been employed for two independent matching procedures with the most obvious choice being the coupling, commonly denoted as $\beta$ (see Eq.~\eqref{def:Wilson}), and the other one being the $\ds$ group characters.
The former option was exploited recently in matching ${\cal O}(\beta^3)$ contributions between $SU(3)$ and $\ds$ through a group decimation technique~\cite{Ji:2020kjk}. These systematic results gave insight into how the Wilson action should be modified when replacing $G$ by $H$.
In this paper, we examine the latter possibility, namely, matching the character contributions between $SU(3)$ and $\ds$.
%

%

The paper is organized as follows.
In Sec.~\ref{sec:char_expansion_1080}, we summarize the group properties of $\ds$, the largest crystallike subgroup of $SU$(3).
This is followed by Sec.~\ref{sec:Cexp}, where we demonstrate the matching via a systematic character expansion. 
Sec.~\ref{sec:res} is reserved for numerical results with analysis of the viability of character expansion.
The paper is finally concluded in Sec.~\ref{sec:concl}.
%

\section{Characters of $\ds$}
\label{sec:char_expansion_1080}

The 1080 elements of the $SU(3)$ subgroup $\ds$ can be classified into $17$ conjugacy classes \cite{Flyvbjerg:1985ad}, generated by its total 17 independent characters for which we denote as $\chi_r^\prime$ with $r=1,2,\cdots, 17$.
These 17 characters are linearly related to a subset of $SU(3)$ characters $\chi_{(\lambda,\mu)}$ which are organized by two \textit{non-negative} integers\footnote{We note that in Ref.~\cite{Ji:2020kjk}, another notation was adopted for general \textit{SU(N)}  characters labeled by $N$ integers $\{\lambda_1,\lambda_2,\cdots,\lambda_N\}$ with $\lambda_1\geq\lambda_2\geq\cdots\geq\lambda_N$. For the special case of \textit{SU}(3), which is our sole interest throughout this paper, all characters can be denoted by only two \textit{non-negative} integers $(\lambda,\mu)$~\cite{Drouffe:1983fv}. We employ the $(\lambda,\mu)$ convention hereafter.} $\lambda$ and $\mu$,
\begin{align}\label{eq:expansion}
     \chi'_r &= \sum_{(\lambda,\mu)} m_{r, (\lambda,\mu)}\,\chi_{(\lambda,\mu)} \, ,
\end{align}
where $m_{r,(\lambda,\mu)}$ are a set of integers systematically obtainable by matching the character definitions of the two groups. 
The explicit linear relations between $\chi_r^\prime$ and $\chi_{(\lambda,\mu)}$ are given in Table~\ref{tab: s1080-in-terms-of-su3} for $r=1,2,\cdots,9,12,\cdots,17$. 

\begin{table*}[ht!]
\caption{Character expansion of $\ds$ characters in terms of $SU(3)$ characters, and in trace representation.}
\begin{center}
\begin{tabular}
{c | c | c  }
\hline\hline
$\chi'_r$ of $\ds$  & $\sum_{(\lambda,\mu)} m_{r, (\lambda,\mu)}\,\chi_{(\lambda,\mu)}$ & Trace Representation \\
\hline
$\chi'_{1}(u)$ & $\chi_{(0,0)}(u)$ & $1$  \\\hline
$\chi'_{2}(u)$ & $\chi_{(1,0)}(u)$ & $\tr(u)$  \\\hline
$\chi'_{3}(u)$ & $\chi_{(0,1)}(u)$ & $\tr(u^\dag)$  \\\hline
$\chi'_{4}(u)$ & $\chi_{(1,1)}(u)$ & $\tr(u)\tr(u^\dag)-1$  \\\hline
$\chi'_{5}(u)$ & $\chi_{(2,0)}(u)$ & $\frac12\left(\tr^2(u)+\tr(u^2)\right)$  \\\hline
$\chi'_{6}(u)$ & $\chi_{(0,2)}(u)$ & $\frac12\left(\tr^2(u^\dagger)+\tr(u^{\dagger2})\right)$  \\\hline
$\chi'_{7}(u)$ & $\chi_{(2,1)}(u)$ & $\frac12\tr(u^\dagger)\left(\tr^2(u)+\tr(u^2)\right)-\tr(u)$  \\\hline
$\chi'_{8}(u)$ & $\chi_{(1,2)}(u)$ & $\frac12\tr(u)\left(\tr^2(u^\dagger)+\tr(u^{\dagger2})\right)-\tr(u^\dagger)$  \\\hline
$\chi'_{9}(u)$ & $\chi_{(3,0)}(u)$ & $\frac16\left(\tr^3(u)+2\tr(u^3)+3\tr(u)\tr(u^2)\right)$  \\\hline
$\chi'_{10}(u)+\chi'_{11}(u)$ & \begin{tabular}{@{}l@{}}$\chi_{(1,1)}(u)+\chi_{(2,2)}(u)$\\$+\chi_{(3,0)}(u)-\chi_{(4,1)}(u)$\end{tabular} & 
\begin{tabular}{@{}l@{}}
     $-1+\frac{1}{4}\left(\tr^2(u)\tr^2(u^\dagger)+\tr^2(u)\tr(u^{\dagger 2})+\tr^2(u^\dagger)\tr(u^2) +\tr(u^{\dagger 2})\tr(u^2) \right)$  \\
     $-\frac{1}{24}\tr(u^\dagger)\left(\tr^4(u)+6\tr^2(u)\tr(u^2)+3\tr^2(u^2)+8\tr(u)\tr(u^3)+6\tr(u^4)\right)$\\
     $+\frac{1}{3}\left(\tr^3(u) +3\tr(u)\tr(u^2)+2\tr(u^3) \right)$
     
\end{tabular}

\\\hline
$\chi'_{12}(u)$ & \begin{tabular}{@{}l@{}}
$\chi_{(3,0)}(u)+\chi_{(2,2)}(u)$\\$+\chi_{(4,1)}(u)-\chi_{(3,3)}(u)$\end{tabular} & 
\begin{tabular}{@{}l@{}}
$-\tr(u)\tr(u^\dagger)-\frac{1}{4}\left(\tr^2(u^\dagger)+\tr(u^{\dagger 2})\right)\left(\tr^2(u)+\tr(u^2) \right)$\\
$+\frac{1}{4}\left(\tr^2(u)\tr^2(u^\dagger)+\tr^2(u)\tr(u^{\dagger 2})+\tr^2(u^\dagger)\tr(u^2)+\tr(u^2)\tr(u^{\dagger 2}) \right)$\\
$+\frac{1}{36}\left(\tr^3(u^\dagger)+3\tr(u^\dagger)\tr(u^{\dagger2})+2\tr(u^{\dagger3}) \right)\left(\tr^3(u)+3\tr(u)\tr(u^2)+2\tr(u^3) \right)$\\
$+\frac{1}{24}\tr(u^\dagger)\left(\tr^2(u)+6\tr^2(u)\tr(u^2)+3\tr^2(u^2)+8\tr(u)\tr(u^3)+6\tr(u^4) \right)$
\end{tabular}  \\\hline
$\chi'_{13}(u)$ & \begin{tabular}{@{}l@{}}$-\chi_{(1,1)}(u)-2\chi_{(3,0)}(u)$\\$-\chi_{(2,2)}(u)+\chi_{(3,3)}(u)$\end{tabular} & \begin{tabular}{@{}l@{}}
$1-\frac{1}{4}\left(\tr^2(u^\dagger) +\tr(u^{\dagger2}) \right)\left(\tr^2(u)+\tr(u^2) \right)+\frac{1}{3}\left( -\tr^3(u)-3\tr(u)\tr(u^2)-2\tr(u^3)\right)$\\
$+\frac{1}{4}\left( -\tr^2(u)\tr^2(u^\dagger)-\tr^2(u)\tr(u^{\dagger2})-\tr^2(u^\dagger)\tr(u^2)-\tr(u^{\dagger2})\tr(u^2)\right)$\\
$+\frac{1}{36}\left(\tr^3(u^\dagger)+3\tr(u^\dagger)\tr(u^{\dagger2})+2\tr(u^{\dagger3}) \right)\left(\tr^3(u)+3\tr(u)\tr(u^2)+2\tr(u^3) \right)$
\end{tabular} \\\hline
$\chi'_{14}(u)$ & $\chi_{(3,1)}(u)-\chi_{(1,2)}(u)$ & \begin{tabular}{@{}l@{}}
$\frac{1}{6}\left(\tr^3(u)+2\tr(u^3)+3\tr(u^2)\tr(u)\right)\tr(u^\dagger)-\frac{1}{2}\left(\tr^2(u)+\tr(u^2)\right)$\\$-\frac{1}{2}\tr(u)\left(\tr^2(u^\dagger)+\tr(u^{\dagger2})\right)+\tr(u^\dagger)$
\end{tabular}   \\\hline
$\chi'_{15}(u)$ & $\chi_{(1,3)}(u)-\chi_{(2,1)}(u)$ & $(\chi'_{14}(u))^*$
\\\hline
$\chi'_{16}(u)$ & \begin{tabular}{@{}l@{}}$\chi_{(3,2)}(u)-\chi_{(2,1)}(u)$\\$-\chi_{(1,3)}(u)$\end{tabular} & \begin{tabular}{@{}l@{}}
     $-\frac{1}{2}\tr(u^\dag)(\tr^2(u)+\tr(u^2))+\frac{1}{12}(\tr^2(u^\dag)+\tr((u^\dag)^2))(\tr^3(u)+3\tr(u)\tr(u^2)+2\tr(u^3))$\\$\quad-\frac12\tr(u^\dagger)\left(\tr^2(u)+\tr(u^{2})\right)+\tr(u)-\frac16\left(\tr^3(u^\dagger)+2\tr(u^{\dagger3})+3\tr(u^{\dagger2})\tr(u^\dagger)\right)\tr(u)$\\$+\frac12\left(\tr^2(u^\dagger)+\tr(u^{\dagger2})\right)$
\end{tabular}  \\\hline
$\chi'_{17}(u)$ & \begin{tabular}{@{}l@{}}$\chi_{(2,3)}(u)-\chi_{(1,2)}(u)$\\$-\chi_{(3,1)}(u)$\end{tabular} & $(\chi^\prime_{16}(u))^*$  \\\hline
\hline
\end{tabular}
\label{tab: s1080-in-terms-of-su3}
\end{center}
\end{table*}

It is worth nothing that although $\chi'_{10}+\chi'_{11}$ is expressible in terms of $\chi_{(\lambda,\mu)}$ in a simple fashion as given in Table~\ref{tab: s1080-in-terms-of-su3}, individual expressions for $\chi'_{10}$ and $\chi'_{11}$ in terms of $SU(3)$ characters are rather lengthy. 
We note that, in practice, $\chi'_{10}$ and $\chi'_{11}$ always appear together as $\chi'_{10}+\chi'_{11}$ and thus the absence of their expressions individually leads to no issues in our derivations.
%

On the other hand, in order to obtain the action of $\chi'_{10}$ and $\chi'_{11}$ on all $\ds$ elements, we exploit the orthonormal condition of the character representation, allowing us to fix the last elements in the character table in Table~\ref{tab: s1080-character-evaluation}.
We point out that the conjugacy class $C_8$ and $C_9$, both of which are composed of traceless $\ds$ elements, are distinguishable only through these two characters $\chi^\prime_{10}$ and $\chi^\prime_{11}$, as can be read from Table~\ref{tab: s1080-character-evaluation}.

\begin{table*}[ht!]
\caption{Character table for $\ds$~\cite{Flyvbjerg:1985ad}. $\mu_1=(1-\sqrt{5})/2$; $\mu_2=(1+\sqrt{5})/2$, $\omega=(1+i\sqrt{3})/2$,  $\omega^*=(1-i\sqrt{3})/2$. The integer in the second row indicates the number of elements in each class, and the letter behind denotes the cycle induced by the class elements.}
\begin{center}
\begin{tabular}
{c | c | c |c | c | c |c | c | c |c | c | c |c | c | c |c | c | c   }
\hline\hline
 class: & $C_1$  &$C_2$ &$C_3$ &$C_4$  &$C_5$ & $C_6$ &$C_7$ &$C_8$ &$C_9$ &$C_{10}$ &$C_{11}$ & $C_{12}$ &$C_{13}$ &$C_{14}$ &$C_{15}$ &$C_{16}$ &$C_{17}$ \\\hline
 &$1E$  &$72c_5$ &$90c_4$ &$45c_6$  &$45c'_6$ & $72c_{15}$ &$72c'_{15}$ &$120c_3$ &$120c'_3$ &$90c_{12}$ &$90c'_{12}$ & $72c'_5$ &$72c_{15}$ &$72c'_{15}$ &$45c_2$ &$1c''_3$ &$1c'''_3$ \\
\hline
$\chi'_1(u)$& $1$& $1$ & $1$ & $1$ & $1$ & $1$ & $1$ & $1$ & $1$ & $1$ & $1$ & $1$ & $1$ & $1$ & $1$ & $1$ & $1$ \\\hline
$\chi'_2(u)$& $3$& $\mu_2$ & $1$ & $\omega$ & $\omega^*$ & $-\mu_1\omega$ & $-\mu_1\omega^*$ & $0$ & $0$ & -$\omega^*$ & $-\omega$ & $\mu_1$ & $-\mu_2\omega^*$ & $-\mu_2\omega$ & $-1$ & $-3\omega^*$ & $-3\omega$  \\ \hline
$\chi'_3(u)$& $3$ & $\mu_2$ & $1$ & $\omega^*$ & $\omega$ & $-\mu_1\omega^*$ & $-\mu_1\omega$ & $0$ & $0$ & $-\omega$ & $-\omega^*$ & $\mu_1$ & $-\mu_2\omega$ & $-\mu_2\omega^*$ & $-1$ & $-3\omega$ & $-3\omega^*$ \\\hline
$\chi'_4(u)$& $8$ & $\mu_2$ & $0$ & $0$ & $0$ & $\mu_1$ & $\mu_1$ & $-1$ & $-1$ & $0$ & $0$ & $\mu_1$ & $\mu_2$ & $\mu_2$ & $0$ & $8$ & $8$ \\\hline
$\chi'_5(u)$& $6$ & $1$ & $0$ & $-2\omega^*$ & $-2\omega$ & $-\omega^*$ & $-\omega$ & $0$ & $0$ & $0$ & $0$ & $1$ & $-\omega$ & $-\omega^*$ & $2$ & $-6\omega$ & $-6\omega^*$ \\\hline
$\chi'_6(u)$& $1$ & $1$ & $0$ & $-2\omega$ & $-2\omega^*$ & $-\omega$ & $-\omega^*$ & $0$ & $0$ & $0$ & $0$ & $1$ & $-\omega^*$ & $-\omega$ & $2$ & $-6\omega^*$ & $-6\omega$ \\\hline
$\chi'_7(u)$& $15$ & $0$ & $-1$ & $\omega$ & $\omega^*$ & $0$ & $0$ & $0$ & $0$ & $\omega^*$ & $\omega$ & $0$ & $0$ & $0$ & $-1$ & $-15\omega^*$ & $-15\omega$ \\\hline
$\chi'_8(u)$& $15$ & $0$ & $-1$ & $\omega^*$ & $\omega$ & $0$ & $0$ & $0$ & $0$ & $\omega$ & $\omega^*$ & $0$ & $0$ & $0$ & $-1$ & $-15\omega$ & $-15\omega^*$ \\\hline
$\chi'_9(u)$& $10$ & $0$ & $0$ & $-2$ & $-2$ & $0$ & $0$ & $1$ & $1$ & $0$ & $0$ & $0$ & $0$ & $0$ & $-2$ & $10$ & $10$ \\\hline
$\chi'_{10}(u)$& $5$ & $0$ & $-1$ & $1$ & $1$ & $0$ & $0$ & $-1$ & $2$ & $-1$ & $-1$ & $0$ & $0$ & $0$ & $1$ & $5$ & $5$ \\\hline
$\chi'_{11}(u)$& $5$ & $0$ & $-1$ & $1$ & $1$ & $0$ & $0$ & $2$ & $-1$ & $-1$ & $-1$ & $0$ & $0$ & $0$ & $1$ & $5$ & $5$ \\\hline
$\chi'_{12}(u)$& $8$ & $\mu_1$ & $0$ & $0$ & $0$ & $\mu_2$ & $\mu_2$ & $-1$ & $-1$ & $0$ & $0$ & $\mu_2$ & $\mu_1$ & $\mu_1$ & $0$ & $8$ & $8$ \\\hline
$\chi'_{13}(u)$& $9$ & $-1$ & $1$ & $1$ & $1$ & $-1$ & $-1$ & $0$ & $0$ & $1$ & $1$ & $-1$ & $-1$ & $-1$ & $1$ & $9$ & $9$ \\\hline
$\chi'_{14}(u)$& $9$ & $-1$ & $1$ & $-\omega^*$ & $-\omega$ & $\omega^*$ & $\omega$ & $0$ & $0$ & $-\omega$ & $-\omega^*$ & $-1$ & $\omega$ & $\omega^*$ & $1$ & $-9\omega$ & $-9\omega^*$ \\\hline
$\chi'_{15}(u)$& $9$ & $-1$ & $1$ & $-\omega$ & $-\omega^*$ & $\omega$ & $\omega^*$ & $0$ & $0$ & $-\omega^*$ & $-\omega$ & $-1$ & $\omega^*$ & $\omega$ & $1$ & $-9\omega^*$ & $-9\omega$ \\\hline
$\chi'_{16}(u)$& $3$ & $\mu_1$ & $1$ & $\omega$ & $\omega^*$ & $-\mu_2\omega$ & $-\mu_2\omega^*$ & $0$ & $0$ & $-\omega^*$ & $-\omega$ & $\mu_2$ & $-\mu_1\omega^*$ & $-\mu_1\omega$ & $-1$ & $-3\omega^*$ & $-3\omega$ \\\hline
$\chi'_{17}(u)$& $3$ & $\mu_1$ & $1$ & $\omega^*$ & $\omega$ & $-\mu_2\omega^*$ & $-\mu_2\omega$ & $0$ & $0$ & $-\omega$ & $-\omega^*$ & $\mu_2$ & $-\mu_1\omega$ & $-\mu_1\omega^*$ & $-1$ & $-3\omega$ & $-3\omega^*$ \\\hline
\hline
\end{tabular}
\label{tab: s1080-character-evaluation}
\end{center}
\end{table*}

\section{Character expansion of Wilson action}
\label{sec:Cexp}
In this section, we first carry out the character expansion for the Boltzmann weight using the Wilson action to sufficiently high orders in \textit{SU}(3). 
%
The domain of the resulting expansion is then reduced from \textit{SU}(3) to $\ds$ with the help of Eq.~(\ref{eq:expansion}) of which the lowest-weight characters are found in Table~\ref{tab: s1080-in-terms-of-su3}.
Clearly, this step lowers the theory complexity at the cost of approximation errors which are quantifiable through theoretical and/or numerical means (e.g.,~\cite{Ji:2020kjk}).
Finally, we derive the effective action $\widetilde S(u)$ over $\ds$ by matching $\e^{-\widetilde S(u)}$ onto the corresponding character expansion.


We start with the pure gauge Wilson action,
\begin{align}
S(U)=-\sum_{p}\frac{\beta}{N}\Re\Tr(U_p)\, ,\qquad U \in SU(3)\, ,
\label{def:Wilson}
\end{align}
where the summation runs over all plaquettes. $U_p$ denotes a product of gauge links in the adjoint representation of \textit{SU}(3).
The resulting Boltzmann weight, $\e^{-S(U)}$, can be decomposed into the orthonormal $SU(3)$ characters basis,
\begin{align}\label{eqn: beta-expansion}
    \e^{-S(U)} &= \sum_{(\lambda,\mu)}\beta_{(\lambda,\mu)}\chi_{(\lambda,\mu)}(U)\, ,
\end{align}
giving rise to a series of character coupling constants $\beta_{(\lambda,\mu)}$.
The resulting partition function central to the classical lattice calculation can therefore be presented as,
\begin{align}
    Z \equiv \int DU\,\e^{-S(U)} = \sum_{(\lambda,\mu)} \int DU \beta_{(\lambda,\mu)}\chi_{(\lambda,\mu)}(U)\, .
\end{align}
For the Wilson action defined in Eq.~\eqref{def:Wilson}, 
the explicit form of $\beta_{(\lambda, \mu)}$ can be written as an infinite sum in terms of Bessel functions~\cite{Drouffe:1983fv}, 
\begin{align}\label{eqn: beta_r_det}
    &\beta_{(\lambda, \mu)}(\beta)\nonumber  \\
    &= \sum_{n=-\infty}^\infty \det \begin{pmatrix}
    I_{\rho+n}(\frac{\beta}{3}) & I_{\sigma-1+n}(\frac{\beta}{3}) & I_{n-2}(\frac{\beta}{3}) \\
    I_{\rho+1+n}(\frac{\beta}{3}) & I_{\sigma+n}(\frac{\beta}{3}) & I_{n-1}(\frac{\beta}{3}) \\
    I_{\rho+2+n}(\frac{\beta}{3}) & I_{\sigma+1+n}(\frac{\beta}{3}) & I_{n}(\frac{\beta}{3})
    \end{pmatrix},
\end{align}
where $\rho=\lambda+\mu$, $\sigma=\mu$, and $I_k$ is the Bessel function of the first kind \cite{Drouffe:1983fv}. 
Explicit  expressions of $\beta_{(\lambda,\mu)}$ that are relevant for our current study are collected in Table~\ref{tab: beta_expansion} in the form of Taylor series of $\beta$.
In practice, we found taking $|n|\leq 40$ in Eq.~\eqref{eqn: beta_r_det} is sufficient numerically.

\begin{table}
\caption{ Series expansions of $\beta_r$, $x\equiv\beta/6$ up to $\mathcal{O}(x^{8})$.}
\begin{center}
\begin{tabular}
{c | c  }
\hline\hline
$(\lambda,\mu)$& $\beta_{(\lambda,\mu)}$ \\
\hline\\[-4mm]
$(0,0)$ & $1-x^2+\frac{x^4}{2}-\frac{11 x^6}{72}+\frac{91 x^8}{2880}+\cdots$  \\[1mm]

$(1,0)$ & $x+\frac{x^2}{2}-x^3-\frac{3 x^4}{8}+\frac{11 x^5}{24}+\frac{11 x^6}{80}-\frac{91 x^7}{720}-\frac{91
   x^8}{2880}+\cdots$  \\[1mm]
   
$(1,1)$ & $-x^2+x^4-\frac{77 x^6}{180}+\frac{13 x^8}{120}+\cdots$  \\[1mm]
  
$(2,0)$ & $\frac{x^2}{2}+\frac{x^3}{2}-\frac{5 x^4}{12}-\frac{11 x^5}{30}+\frac{x^6}{6}+\frac{56 x^7}{365}-\frac{59 x^8}{1440}+\cdots$  \\[1mm]

$(2,1)$ & $-\frac{x^3}{2}-\frac{5 x^4}{24}+\frac{11 x^5}{24}+\frac{7 x^6}{48}-\frac{13 x^7}{72}-\frac{3 x^8}{64}+\cdots$  \\[1mm]

$(3,0)$ & $\frac{x^3}{6}+\frac{x^4}{4}-\frac{x^5}{12}-\frac{25 x^6}{144}+\frac{x^7}{48}+\frac{113 x^8}{2016}+\cdots$  \\[1mm]

$(2,2)$ & $\frac{x^4}{4}-\frac{9 x^6}{40}+\frac{27 x^8}{320}-\frac{187 x^{10}}{10080}+\hdots$  \\[1mm]
  
$(3,1)$ & $-\frac{x^4}{6}-\frac{2x^5}{15}+\frac{x^6}{8}+\frac{4x^7}{45}-\frac{3 x^8}{70}+\cdots$  \\[1mm]
   
$(4,0)$ & $\frac{x^4}{24}+\frac{x^5}{12}-\frac{53 x^7}{1008}-\frac{19 x^8}{2880}+\cdots$ \\[1mm]

$(3,2)$ & $\frac{x^5}{12}+\frac{7 x^6}{240}-\frac{49 x^7}{720}-\frac{3 x^8}{160}
   +\cdots$  \\[1mm]
   
$(4,1)$ & $-\frac{x^5}{24}-\frac{7 x^6}{144}+\frac{x^7}{48}+\frac{29 x^8}{960}+\cdots$  \\[1mm]

$(5,0)$ & $\frac{x^5}{120}+\frac{x^6}{48}+\frac{x^7}{180}-\frac{13 x^8}{1152}+\cdots$  \\[1mm]

$(3,3)$ & $-\frac{x^6}{36}+\frac{x^8}{45}+\cdots$  \\[1mm]
   
$(4,2)$ & $\frac{x^6}{48}+\frac{x^7}{72}-\frac{71 x^8}{5040}+\cdots$  \\[1mm]

$(5,1)$ & $-\frac{x^6}{120}-\frac{4x^7}{315}+\frac{x^8}{720}+\cdots$  \\[1mm] 

$(6,0)$ & $\frac{x^6}{720}+\frac{x^7}{240}+\frac{x^8}{480}+\cdots$  \\[1mm]

\hline
\end{tabular}
\label{tab: beta_expansion}
\end{center}
\end{table}
%

We are now ready to extract the effective action over $\ds$, the finite discrete crystal subgroup of \textit{SU}(3). 
The starting point of our procedure is to substitute $U\in SU(3)$ on the R.H.S. of Eq.~\eqref{eqn: beta-expansion} by $u\in\ds$,  effectively reducing the domain of the character expansion.  
The $SU(3)$ characters $\chi_{(\lambda,\mu)}(U)$ are then mapped to their $\ds$ counterparts $\chi^\prime_r(u)$ using either Eq.~\eqref{eq:expansion}, or Table~\ref{tab: s1080-character-evaluation} by solving a system of 17 linear equations.
Subsequently, the \textit{SU}(3) character couplings $\beta_{(\lambda,\mu)}$ are also rearranged into $\ds$ character couplings $\beta^\prime_r$. 
Detailed relations between $\beta^\prime_{r}$ and $\beta_{(\lambda,\mu)}$ are collected in Table~\ref{tab: beta_prime_in_beta}.
The resulting $\ds$ character expansion is finally matched onto the $\ds$ effective action $\widetilde S(u)$ as follows,
\begin{align}\label{eqn: s1080-character-expansion}
    e^{-\widetilde S(u)}=\sum_{r=1}^{17} \beta'_r\chi'_r(u)\,, \qquad u\in \ds\, .
\end{align}
%
We emphasize that once the order of truncation for the \textit{SU}(3) character expansion is determined, the R.H.S of Eq.~\eqref{eqn: s1080-character-expansion} is completely fixed through Eq.~\eqref{eq:expansion} (or Table~\ref{tab: s1080-in-terms-of-su3}). 
The relations between $\beta^\prime_r$ and $\beta_{(\lambda,\mu)}$ given in Table~\ref{tab: beta_prime_in_beta} provide a complete basis for \textit{SU}(3) character $\chi_{(\lambda,\mu)}$ expansion up to $0\leq\lambda+\mu\leq 6$. It is also important to remark that here we have systematically neglected the contribution of $SU(3)$ elements which are absent in $\ds$, effectively providing us with a ``leading-order" approximation. Such approximations are improvable systematically by parametrizing $U =\epsilon u$~\cite{Ji:2020kjk} yielding,
\begin{align}
    Z &=\sum_{u\in\ds} \sum_{(\lambda,\mu)} \int D\epsilon\,  \beta_{(\lambda,\mu)}\chi_{(\lambda,\mu)}(\epsilon u)\notag\\
    &=\sum_{u\in\ds} \sum_{(\lambda,\mu)} \beta_{(\lambda,\mu)}\, F(V_{(\lambda,\mu)}, \chi_{(\lambda,\mu)}(u))\, ,\label{eq:Zint}
\end{align}
where 
\begin{align}\label{eq: F}
    \int D\epsilon\,\chi_{(\lambda,\mu)}(\epsilon u) = F(V_{(\lambda,\mu)}, \chi_{(\lambda,\mu)}(u))=\widetilde{F}(V_r^\prime, \chi_r^\prime(u))\, .
\end{align}
We have schematically applied relations in Table.~\ref{tab: s1080-in-terms-of-su3} or~\ref{tab: s1080-character-evaluation} at the second equality in Eq.~\eqref{eq: F}, and $\displaystyle V_{(\lambda,\mu)}=\frac{1}{d_{(\lambda,\mu)}}\int_\Omega D\epsilon \Re{\chi_{(\lambda,\mu)}}(\epsilon)$, where $d_{(\lambda,\mu)}$ is the dimension of the character $\chi_{(\lambda,\mu)}$ as defined in~\cite{Drouffe:1983fv}. The function $F(V_{(\lambda,\mu)},\chi_{(\lambda,\mu)})$, which is linear in $\chi_{(\lambda,\mu)}$, is obtainable through systematic albeit tedious algebraic manipulations as explicitly demonstrated in~\cite{Ji:2020kjk}. 
At higher orders, however, it is more advantageous to derive the $\chi_r^\prime(u)$ dependence of $\widetilde F(V_r^\prime, \chi_r^\prime(u))$ numerically by solving a system of 17 linear equations, generated by $u\in\ds$ from 17 different conjugacy classes.
The domain of Eq.~\eqref{eq:Zint} hence reduces to $\ds$ leading to 
\begin{equation}
    \displaystyle \chi_{(\lambda,\mu)} = \sum_{r=1}^{17} c_{(\lambda,\mu)}^r\chi^\prime_{r}\, .
\end{equation}

\begin{table}
\caption{$\beta'_r$ as linear combinations of $\beta_{(\lambda,\mu)}$.}
\begin{center}
\begin{tabular}
{c | c }
\hline\hline
$\beta'_1$& $\beta_{(0,0)}+2\beta_{(6,0)}$\\
\hline
$\beta'_2$& $\beta_{(1,0)}+\beta_{(5,0)}+\beta_{(5,1)}$\\
\hline
$\beta'_3$& $\beta'_2$\\
\hline
$\beta'_4$& $\beta_{(1,1)}+2\beta_{(4,1)}+\beta_{(3,3)}+2\beta_{(6,0)}$\\
\hline
$\beta'_5$& $\beta_{(2,0)}+\beta_{(4,0)}+2\beta_{(4,2)}+\beta_{(5,1)}$\\
\hline
$\beta'_6$& $\beta'_5$\\
\hline
$\beta'_7$& $\beta_{(2,1)}+\beta_{(3,1)}+2\beta_{(3,2)}+\beta_{(5,0)}+2\beta_{(4,2)}+2\beta_{(5,1)}$\\
\hline
$\beta'_8$& $\beta'_7$\\
\hline
$\beta'_9$& $2\beta_{(3,0)}+2\beta_{(4,1)}+2\beta_{(3,3)}$\\
\hline
$\beta'_{10}$& $\beta_{(2,2)}+\beta_{(3,3)}+2\beta_{(6,0)}$\\
\hline
$\beta'_{11}$& $\beta'_{10}$\\
\hline
$\beta'_{12}$& $\beta_{(2,2)}+2\beta_{(4,1)}+\beta_{(3,3)}$\\
\hline
$\beta'_{13}$& $\beta_{(2,2)}+2\beta_{(4,1)}+2\beta_{(3,3)}+2\beta_{(6,0)}$\\
\hline
$\beta'_{14}$& $\beta_{(3,1)}+\beta_{(4,0)}+\beta_{(3,2)}+2\beta_{(4,2)}+\beta_{(5,1)}$\\
\hline
$\beta'_{15}$& $\beta'_{14}$\\
\hline
$\beta'_{16}$& $\beta_{(3,2)}+\beta_{(5,0)}$\\
\hline
$\beta'_{17}$& $\beta'_{16}$\\
\hline

\end{tabular}
\label{tab: beta_prime_in_beta}
\end{center}
\end{table}
With the determined values of $\beta'_i$, we can now reexponentiate to obtain the effective action $\widetilde S(u)$ in Eq.~(\ref{eqn: s1080-character-expansion}).
%
Rewriting the effective action $\widetilde S(u)$ as an expansion in terms of $\chi_r^\prime$, we obtain  
the following matching formula from which the coefficients $\gamma_i$ are fixed to the targeted order\footnote{Both $\beta^\prime_i$ and $\gamma_i$ are $\beta$-dependent.} in $\beta$,
\begin{align}
  \exp(-\widetilde S(u)) =   \exp{\left(-\sum_{i=1}^{17}\gamma_i\chi'_i\right)} &= \sum_{i=1}^{17}\beta'_i\chi'_i\, ,
\end{align}
which implies
\begin{align}
    -\sum_{i=1}^{17}\gamma_i\chi'_i &= \log \left(1+\sum_{i=1}^{17}\beta'_i\chi'_i-1\right)\equiv \log\left(1+z\right).\label{eqn:re-exp}
\end{align}
Coefficients $\gamma_i$ are then extracted by expanding the R.H.S. with respect to $z$ and matching onto the L.H.S. in terms of the 17 characters, the completeness of $\chi_r^\prime$ as a character representation of $\ds$ validates  
\begin{align}\label{eqn: coeff_c_chi}
    \chi^\prime_i\chi^\prime_j &= \sum_{k=1}^{17} c_{ijk}\chi'_k\, ,
\end{align}
with $c_{ijk}$ being integers easily calculable from Table~\ref{tab: s1080-character-evaluation} by solving a system of 17 linear equations with 17 unknowns.

Substituting Eq.~\eqref{eqn: coeff_c_chi} into Eq.~\eqref{eqn:re-exp} then yields all the coefficients $\gamma_i$ as polynomials of the strong coupling $\beta$. Explicit expressions with parametrization invariance, a general principle for constructing effective actions which we explain below, are summarized in Table~\ref{tab: final_gamma_fifth_order_invariance}. 

Finally, it is important to notice that the original Wilson action $S(U)$ is invariant under reparameterization $\{\beta,U\}\mapsto\{\beta/c, cU\}$ with $c\neq0$ being an arbitrary constant. This property, which we call reparemeterization invariance (RI), is no longer present when the character expansion is truncated at finite orders. We explictly reintroduce it to remove finite order terms in the effective action $\widetilde S(u)$ that are destined to sum up to zero. This not only gives us a much simpler expression for the effective action but also helps reduce the computational cost in lattice simulations and should improve agreement with the $SU(3)$ action.

\begin{table}[ht!]
\caption{$\gamma_i$ computed in terms of $\beta$ to $\mathcal{O}(\beta^5)$ with reparameterization invariance.}
\begin{center}
\begin{tabular}
{c | c  }
\hline\hline
$\gamma_1$& \begin{tabular}{@{}l@{}}
$-\frac{1}{18}\beta ^2-0.00308642 \beta ^4$
\end{tabular}\\
\hline

$\gamma_2$& \begin{tabular}{@{}l@{}}
$\frac{1}{6}\beta +0.00925926 \beta ^3+0.00174683 \beta ^5$
\end{tabular}\\
\hline

$\gamma_3$& \begin{tabular}{@{}l@{}}
$\gamma_2$
\end{tabular}\\
\hline

$\gamma_4$& \begin{tabular}{@{}l@{}}
$-\frac{1}{18} \beta ^2-0.00308642 \beta ^4-0.000364369 \beta ^5$
\end{tabular}\\
\hline

$\gamma_5$& \begin{tabular}{@{}l@{}}
$-0.000257202 \beta ^4$
\end{tabular}\\
\hline

$\gamma_6$& \begin{tabular}{@{}l@{}}
$\gamma_5$
\end{tabular}\\
\hline

$\gamma_7$& \begin{tabular}{@{}l@{}}
$0.00925926 \beta ^3-0.000257202 \beta ^4+0.00524049 \beta ^5$
\end{tabular}\\
\hline

$\gamma_8$& \begin{tabular}{@{}l@{}}
$\gamma_7$
\end{tabular}\\
\hline

$\gamma_9$& \begin{tabular}{@{}l@{}}
$-0.000728738 \beta ^5$
\end{tabular}\\
\hline

$\gamma_{10}$& \begin{tabular}{@{}l@{}}
$-0.00308642 \beta ^4$
\end{tabular}\\
\hline

$\gamma_{11}$& \begin{tabular}{@{}l@{}}
$\gamma_{10}$
\end{tabular}\\
\hline

$\gamma_{12}$& \begin{tabular}{@{}l@{}}
$-0.00308642 \beta ^4-0.000364369 \beta ^5$
\end{tabular}\\
\hline

$\gamma_{13}$& \begin{tabular}{@{}l@{}}
$\gamma_{12}$
\end{tabular}\\
\hline

$\gamma_{14}$& \begin{tabular}{@{}l@{}}
$-0.000257202 \beta ^4+0.00174683 \beta ^5$
\end{tabular}\\
\hline

$\gamma_{15}$& \begin{tabular}{@{}l@{}}
$\gamma_{14}$
\end{tabular}\\
\hline

$\gamma_{16}$& \begin{tabular}{@{}l@{}}
$0.00174683 \beta ^5$
\end{tabular}\\
\hline

$\gamma_{17}$& \begin{tabular}{@{}l@{}}
$\gamma_{16}$
\end{tabular}\\
\hline
\end{tabular}
\label{tab: final_gamma_fifth_order_invariance}
\end{center}
\end{table}

The effective action we have derived from the character expansion in terms of the $\gamma_i$ in Table~\ref{tab: final_gamma_fifth_order_invariance} can be compared to the group decimated action, $S_{GD}$, defined by the couplings in Table III of~\cite{Ji:2020kjk}. $S_{GD}$ was only calculated to $\mathcal{O}(\beta^3)$ and thus the scaling behavior for certain $\chi'_r$ are unknown.  In contrast, using the character expansion we can determine the leading power of $\beta$ for all $\chi'_r$.  

Between the two schemes, there are different  $\beta$ dependences.  In particular, $\chi'_1$ in the character expansion contains only odd powers of $\beta$ while $S_{GD}$ generates all new terms at all orders.  When comparing the two ad-hoc modified $\ds$ actions which individually added $\chi'_4$~\cite{lammtocome} and $\chi'_5$~\cite{Alexandru:2019nsa,Alexandru:2021jpm}, the group decimation procedure suggested no preference -- they both are generated at $\mathcal{O}(\beta^2)$. Within the character expansion, only $\chi'_4$ appears at $\mathcal{O}(\beta^2)$.  $\chi'_5$ only arises at $\mathcal{O}(\beta^4)$.
\section{Numerical Results}
\label{sec:res}

\begin{figure}[ht!]
  \centering
  \includegraphics[width=\linewidth]{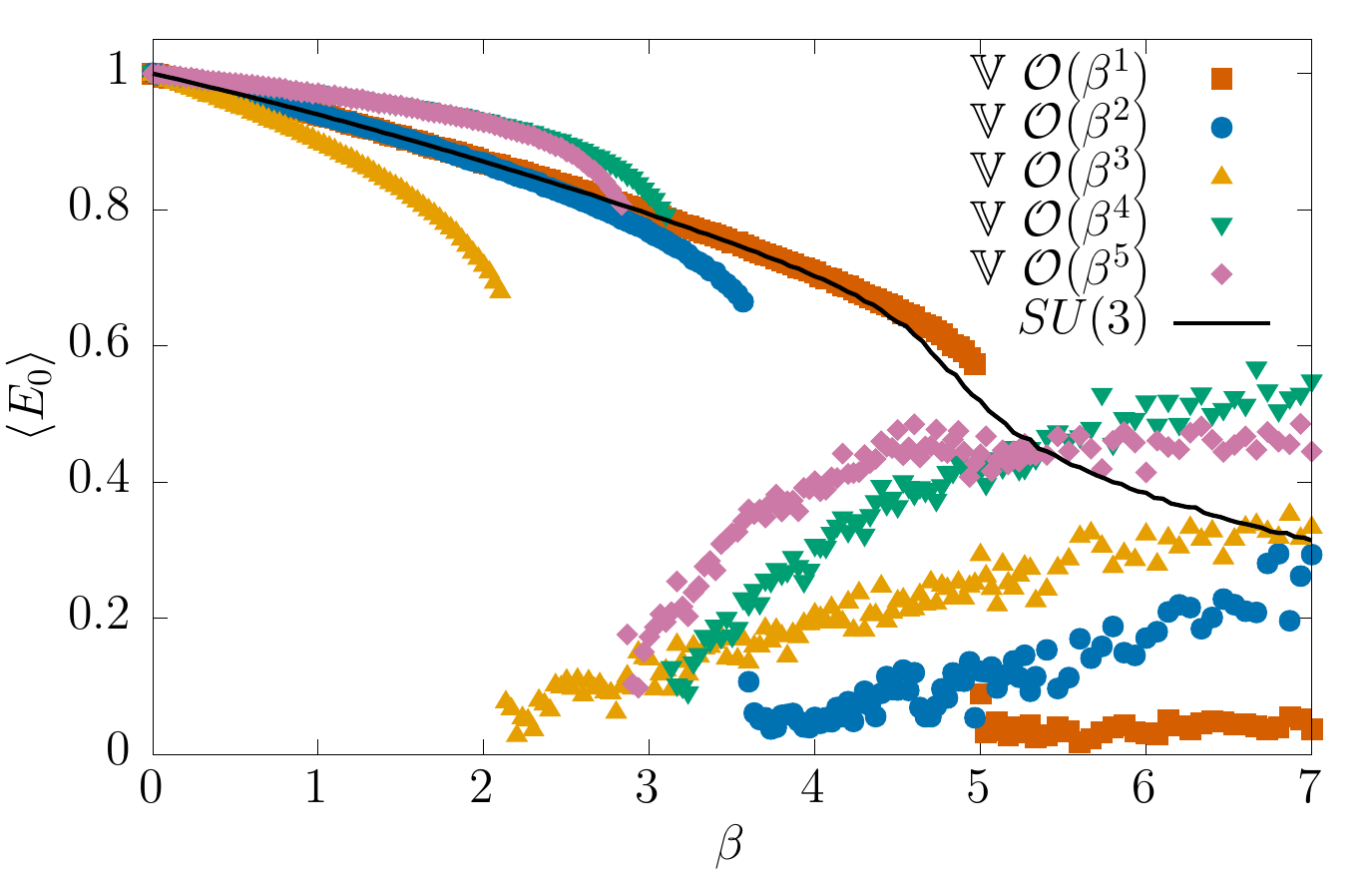}
  \caption{Average energy per plaquette, $\langle E_0\rangle = 1-\Re\langle\Tr u_p\rangle /3$, vs $\beta$ on a $4^4$ lattice for $\ds$ action with corrections of: (\protect\redsq) $\mathcal{O}(\beta^1)$,  (\protect\blcir) $\mathcal{O}(\beta^2)$, (\protect\ortri) $\mathcal{O}(\beta^3)$, (\protect\grtri) $\mathcal{O}(\beta^4)$, and (\protect\pudia) $\mathcal{O}(\beta^5)$. The black line is the $SU(3)$ result.}
  \label{fig:plaqv}
\end{figure}

In order to gauge the effectiveness of our approximations, the effective $\ds$ action induced by character expansion was simulated at each order in $\beta$.  For these computations 10$^2$ configurations separated by 10$^2$ sweeps were collected on a $4^4$ lattice. The average energy per plaquette $\langle E_0\rangle$ versus $\beta$ is plotted in Fig.~\ref{fig:plaqv}.

If the effective action behaved as the continuous group action, $\langle E_0\rangle$ would be observed to be monotonic in $\beta$. We observe that including $\mathcal{O}(\beta^2)$ or higher terms that instead of the simple freeze-out story of discrete groups, after a large drop in $\langle E_0\rangle$ the expectation value begins to rise again before asymptoting to a fixed value. Comparing the actions at increasing orders in $\beta$, the convergence to the continuous group results is slow and alternating. Despite this, higher-order effective action seems to correspond to a wider range of $\langle E_0\rangle$ being accessible.  For the $\mathcal{O}(\beta^5)$ action, it was found to be possible to obtain $0.2\leq \langle E_0\rangle\leq 0.4$. This naively suggests that the action might allow for smaller lattice spacing than the unmodified one - including physics within the scaling regime. 

It should be noted that the behavior observed by this expansion is qualitatively different to that of the group decimation procedure of \cite{Ji:2020kjk}.  In that work, the effective actions obtained at various orders of $\beta$ were found to lack the first-order phase-transition behavior of a discrete group, but the range of $\langle E_0\rangle$ was limited to much larger values of $0.5-1$  in the small $\beta$ region. 

\section{Conclusion}
\label{sec:concl}
In this work, we used the character expansion to develop a systematic method for improving lattice actions that replace continuous gauge group $SU(3)$ by its discrete subgroup $\ds$. Moreover, this method can be generalized to any continuous gauge group $SU(N)$. We also spell out a new principle called reparameterization invariance as a guideline for constructing effective actions, allowing to reduce the cost of computation and match the original action more closely at the same time. This is the ongoing effort toward developing efficient digitization schemes on quantum computers. 

 We computed to $\mathcal{O}(\beta^5)$ for the single-plaquette $\ds$ action as an approximation of the Wilson action of $SU(3)$. 
 These higher-order terms suggest a different scaling with $\beta$ compared to previous expansions based on group decimation \cite{Ji:2020kjk}. 
The higher-order (${\cal O}(\beta^2)$) result in this work does not take into account the contribution of quantum fluctuations around $\ds$ elements. So as expected it deviates from the $SU(3)$ result with a hard truncation in the character expansion. It would be beneficial to include the quantum fluctuation through Table~\ref{tab: s1080-in-terms-of-su3} in future work to achieve better approximations.
 
Moreover, while the group decimation procedure was observed to suggest the necessity of introducing two characters at $\mathcal{O}(\beta^2)$, the character expansion only require the adjoint character to be present.

 The most immediate next steps are to compute more Euclidean observables (i.e. Wilson flow parameter and pseudocritical temperature and etc.) from the character expansion action to determine the lattice spacings achievable and to compare them to~\cite{Alexandru:2019nsa,Alexandru:2021jpm}. Given the large corrections order by order for $\ds$ and the interesting even versus odd behavior, additional work should be devoted to computing the higher order contributions and corrections induced by fluctuations around $\ds$ elements - potentially via some resummation technique.

Another obvious step in studying the feasibility of this procedure is to derive the modified Hamiltonian, and explicitly construct the primitive quantum gates \`a la \cite{Lamm:2019bik,Alam:2021uuq}. Together with classical lattice results, this would allow for complete resource counts for extracting continuum physics.

\begin{acknowledgments}
The authors would like to thank Justin Thaler for helpful comments on this work. Y.J. is grateful for the support of Deutsche Forschungsgemeinschaft
(DFG, German Research Foundation) grant SFB TR 110/2. H.L is supported by the Department of Energy through the Fermilab QuantiSED program in the area of ``Intersections of QIS and Theoretical Particle Physics". Fermilab is operated by Fermi Research Alliance, LLC under contract number DE-AC02-07CH11359 with the United States Department of Energy. S.Z. is supported by the National Science Foundation CAREER award (grant CCF-1845125).
\end{acknowledgments}


\bibliographystyle{apsrev4-1}
\bibliography{wise}

\end{document}